\documentclass{amsart}

\title{A rearrangement step with potential uses in priority queues}
\author{Boris Alexeev \and M. Brian Jacokes}

\usepackage{fullpage, setspace, xspace, pstricks}
\begin{document}
\begin{abstract}
  Link-based data structures, such as linked lists and binary search
  trees, have many well-known rearrangement steps allowing for efficient
  implementations of insertion, deletion, and other operations.  We
  describe a rearrangement primitive designed for link-based,
  heap-ordered priority queues in the comparison model, such as those
  similar to Fibonacci heaps or binomial heaps.

  In its most basic form, the primitive rearranges a collection of
  heap-ordered perfect binary trees.  Doing so offers a data structure
  control on the number of trees involved in such a collection, in
  particular keeping this number logarithmic in the number of elements.
  The rearrangement step is free from an amortized complexity standpoint
  (using an appropriate potential function).
\end{abstract}
\date{\today}
\maketitle

\renewcommand{\P}        {\ensuremath{\mathcal{P}}}
\renewcommand{\phi}      {\ensuremath{\varphi}}

\newcommand{\op}      [1]{\ensuremath{\textsf{#1}}\xspace}
\newcommand{\insertkey}  {\op{insert}}
\newcommand{\meld}       {\op{meld}}
\newcommand{\deletemin}  {\op{delete-min}}
\newcommand{\decreasekey}{\op{decrease-key}}
\newcommand{\findmin}    {\op{find-min}}
\newcommand{\delete}     {\op{delete}}
\newcommand{\updatekey}  {\op{update-key}}
\newcommand{\increasekey}{\op{increase-key}}
\newcommand{\makequeue}  {\op{make-queue}}

\newcommand{\tamo}       {\ensuremath{t_{\text{amortized}}}}
\newcommand{\tact}       {\ensuremath{t_{\text{actual}}}}

\section{Introduction}

The priority queue, or \emph{heap}, is perhaps the simplest data
structure beyond stacks and FIFO queues.  Originally designed as part of
the sorting algorithm \emph{heapsort}~\cite{w}, heaps are now used in a
variety of algorithms, particularly graph-theoretic ones (such as
Dijkstra's single-source shortest-paths algorithm).  In its most basic
form, a heap represents a collection $\P$ of elements from a
totally-ordered set, together with the operations \insertkey and
\deletemin; the latter returns the name of the least element after
deleting it from~$\P$.  More useful heap implementations will include
the operations \decreasekey, \meld, \delete, and \makequeue, which we do
not describe in detail here.

The importance and utility of heaps stems from fast access to the
current minimum element, unlike the fixed access patterns of stacks and
queues.  Simultaneously, the absence of operations like iteration
through $\P$ in sorted order allows heaps to answer queries faster than
full-fledged binary search trees.  Many heap implementations have been
developed, including the original binary heap~\cite{w}, binomial
heap~\cite{v}, and Fibonacci heaps~\cite{ft}.  The basic atom of these
implementations is the concept of a \emph{heap-ordered tree}, which is a
rooted tree such that each element is no greater than any of its
children.  In such a tree, clearly the smallest element is the root.

Most implementations of heaps in the tradition of binomial heaps
maintain a collection of heap-ordered trees represented explicitly, that
is, with pointers.  (An important exception is binary heaps which
maintain a single heap-ordered tree represented implicitly in an array.)
In order to perform updates and answer queries efficiently, there are a
few primitive operations that rearrange elements in a manner preserving
the heap-ordering.  Two of the most important are \emph{bubble-up} (also
known as up-heap) and \emph{bubble-down} (also known as down-heap),
which address local inconsistencies in the heap-ordering at a leaf or
root, respectively.  There are also other primitives, such as the
procedure for merging two binomial trees.

We introduce a new rearrangement primitive that rearranges a collection
of heap-ordered trees.  In short, given three perfect heap-ordered
binary trees, the rearrangement creates one larger perfect heap-ordered
tree and two smaller ones.

\section{Rearrangement step}

Recall that a binary tree is \emph{perfect} (also called \emph{full}) if
all leaves are at the same depth.  Such a tree will necessarily contain a
number of nodes one less than a power of two.

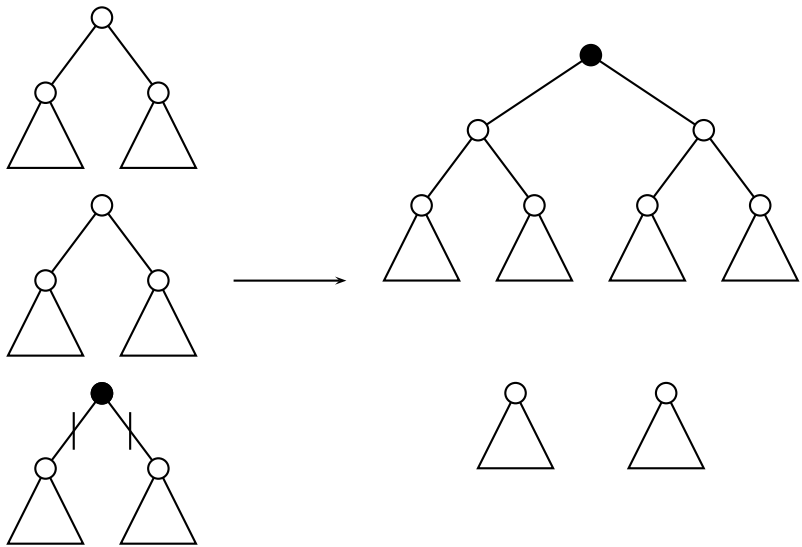
\begin{figure}[ht]
  \begin{pspicture}(-0.25,-0.25)(10.25,7.25)
    \multirput(0,0)(0,2.5){3}{
      \pspolygon(0,0)(1,0)(0.5,1)
      \pspolygon(1.5,0)(2.5,0)(2,1)
      \psline(0.5,1)(1.25,2)(2,1)
      \pscircle[fillcolor=white,fillstyle=solid](0.5,1){0.15}
      \pscircle[fillcolor=white,fillstyle=solid](2,1){0.15}
      \pscircle[fillcolor=white,fillstyle=solid](1.25,2){0.15}
    }
    \pscircle[fillcolor=black,fillstyle=solid](1.25,2){0.15}
    \psline(0.875,1.25)(0.875,1.75)
    \psline(1.625,1.25)(1.625,1.75)
    \psline{->}(3,3.5)(4.5,3.5)

    \psline(6.25,5.5)(7.75,6.5)(9.25,5.5)
    \pscircle[fillcolor=black,fillstyle=solid](7.75,6.5){0.15}
    \multirput(5,3.5)(3,0){2}{
      \pspolygon(0,0)(1,0)(0.5,1)
      \pspolygon(1.5,0)(2.5,0)(2,1)
      \psline(0.5,1)(1.25,2)(2,1)
      \pscircle[fillcolor=white,fillstyle=solid](0.5,1){0.15}
      \pscircle[fillcolor=white,fillstyle=solid](2,1){0.15}
      \pscircle[fillcolor=white,fillstyle=solid](1.25,2){0.15}
    }

    \multirput(6.25,1)(2,0){2}{
      \pspolygon(0,0)(1,0)(0.5,1)
      \pscircle[fillcolor=white,fillstyle=solid](0.5,1){0.15}
    }
    
  \end{pspicture}
  \caption{A rearrangement step: three trees of height $h$ are rearranged to make one tree of
    height $h+1$ and two trees of height $h-1$ by selecting the smallest root (shaded) and
    rearranging children.}
  \label{figure:rearrangement}
\end{figure}

A \emph{rearrangement step} is a method of taking three perfect binary
trees of the same size (say, of height $h$) and producing one perfect
binary tree of height $h+1$ and two perfect binary trees of height
$h-1$, in constant time.  This can be accomplished by comparing the
three roots, selecting the smallest one, removing it from its tree
(leaving two subtrees), and making it the root of a new larger tree with
the other two trees as children.  See Figure~\ref{figure:rearrangement}
for visual accompaniment.

Suppose a heap is implemented by maintaining a collection of perfect,
heap-ordered trees~\cite{ss, bsg}.  Using the rearrangement step
iteratively if necessary, the data structure may ensure that no more
than two trees of any given height are present in the collection.
Because the number of elements in each tree grows exponentially with the
height, this ensures that no more than logarithmically-many trees are in
the collection overall.  Other operations, such as \deletemin or \meld,
are thus easier to implement in logarithmic time.

From the point of view of amortized complexity~\cite{t}, the
rearrangement step pays for itself.  More specifically, consider a
\emph{potential function} $\phi$ that is the sum of the heights of the
trees in a collection of perfect binary trees.  Evidently, this
potential function goes down by one every rearrangement step.
Therefore, the amortized time complexity $\tamo = \tact + \phi$
(normalized so that each rearrangement step takes one unit of $\tact$)
does not change when a rearrangement step is performed.  Since in any
sequence of operations beginning with an empty queue, the initial value
of $\phi$ (namely, zero) is at most the final value of $\phi$, the
amortized time $\tamo$ is an upper bound on the actual time $\tact$.  Of
course, for this analysis to be useful, the other operations must play
well with the potential function.  In our case, straightforward
implementations of the heap operations do not modify $\phi$ by much.

Of course, the rearrangement step need not be limited in its application
to perfect binary trees only.  However, its implementation in that case
is most straightforward so we limit our description to that situation.

\section{Applications}

The rearrangement step described in this paper was also discovered by
Claus Jensen, independently of these authors.  Furthermore, Elmasry,
Jensen, and Katajainen have used the step to implement a priority queue
that guarantees $O(1)$ worst-case cost per \insertkey and $O(\log n)$
worst-case cost per \delete, where $n$ is the number of elements
stored.~\cite{ejkc, ejkj}

More specifically, those authors analyze a number system based on powers
of two minus one (that is, the size of a perfect heap).  A
straightforward use of the rearrangement step would then allow each
digit to be at most two; the \emph{skew binary number system} has this
property.  However, in order to guarantee \emph{worst-case} times rather
than simply amortized times, the authors allow each digit to be at most
four.  Our writeup complements that of Elmasry, Jensen, and Katajainen
by isolating the rearrangement primitive, which is used inline in their
paper under the name \emph{fix}.

Thus this rearrangement step, in addition to being a theoretical
addition to the basic link mechanisms used in the existing plethora of
link-based priority queues, has already found applications.

\bibliographystyle{amsalpha}
\providecommand{\bysame}{\leavevmode\hbox to3em{\hrulefill}\thinspace}
\providecommand{\MR}{\relax\ifhmode\unskip\space\fi MR }
\providecommand{\MRhref}[2]{%
  \href{http://www.ams.org/mathscinet-getitem?mr=#1}{#2}
}
\providecommand{\href}[2]{#2}

\end{document}